\documentclass{article}

\usepackage[final]{nips_2018}

\usepackage[utf8]{inputenc} 
\usepackage[T1]{fontenc}    
\usepackage{hyperref}       
\usepackage{url}            
\usepackage{booktabs}       
\usepackage{amsfonts}       
\usepackage{nicefrac}       
\usepackage{microtype}      
\usepackage{graphicx}
\usepackage{rotating}
\usepackage[export]{adjustbox}[2011/08/13]
\title{Statistical Impact of New York Health Legislation}

\author{
 Will Long\And Joe Zhou \And Zhirui Hu \And Yuntian Deng \And \vspace{-7mm} \\
 $\texttt{\small \{wlong@college, jzhou02@g, zhiruihu@g, dengyuntian@seas\}.harvard.edu} $ \vspace{4mm} \\
Harvard University \\
Cambridge, MA, USA
}

\begin{document}

\maketitle

\begin{abstract}
As the US Government plays an increasing role in health care, it becomes essential to understand the impact of expensive legislation on actual outcomes. New York, having spent the last decade heavily legislating health-related behavior, represents a unique test case to gain insight about what factors cause health care legislation to succeed or fail. We present a longitudinal study comparing bills across 13 Health Areas to measure the effect legislation in that Area had on 311 hotline service complaints. We find that there is statistically significant evidence with p-value $p=0.05$ that legislation in the Hazardous Materials Health Area correlated with a positive change in outcomes. The other Health Areas correlated with changes, but were not statistically significant.

\end{abstract}

\section{Introduction}
State Legislatures pass over 120,000 bills each year totalling over 3 million words [1]. The time from introduction to enactment to agency implementation takes an average of 354 days [2, 3]. The actual effect of these bills in the real-world is long-forgotten by the time they become measurable.

Informed law-making is uniquely important to achieving good health outcomes in the US. Previous studies have shown that state-level legislation can boost health care utilization rates by over $90\%$ and even inspire individuals from neighboring states to live more healthily [4].

New York offers a compelling case study for understanding the effects of legislation on health outcomes. Mayor Michael Bloomberg's "Health in all Policies" legislation during his 2002-2013 tenure tackled air pollution, dieting, physical activity, and smoking and included such infamous policies as the fountain drink size limiting and curb smoking prohibition. We propose the following research question: 

\textit{How effective is New York state health legislation at producing better health outcomes and how can they be better formulated to achieve those ends?}

\section{Data}
\subsection{Datasets: Service Requests \& NY Bills}
For our study, we utilize the publicly-available dataset \textit{311\_service\_requests}, a New York state compilation of 1 million service requests made to the 311 reporting hotline from 2010-2018, as a self-reported metric of health outcome. These complaints are divided according to several health categories documented in Figure 1 (e.g. Air Quality, Hazardous Materials, Rodent, etc.)

To correlate these outcomes with legislative activity, we also utilize a publicly-available dataset \textit{nys\_bills} of $91,000$ New York State Legislature bills passed between 2011-2016. Relevant columns are displayed in Table 1.
\begin{figure}[!t]
    \centering
    \includegraphics[width=0.7\textwidth]{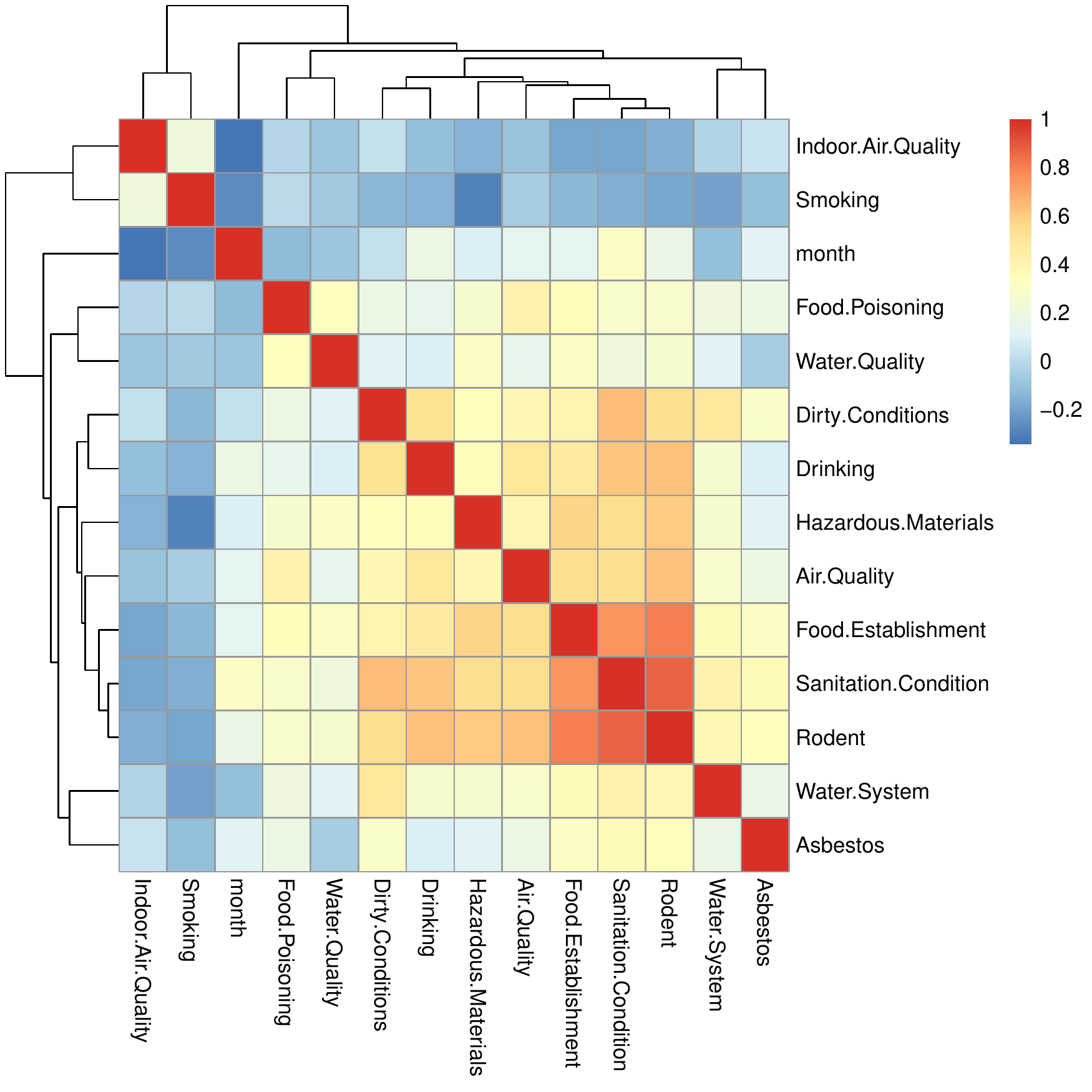}
    \caption{Correlation between Health Areas. Note the $3 \times 3$ red block formed by \textit{Food Establishment}, \textit{Sanitation Condition} and \textit{Rodent} that indicates a relationship between the three. Consequently, we should see legislative outcome trends between correlated Health Areas move in tandem.}
    \label{fig:correlation_between_categories}
\end{figure}
\begin{table}[ht]
\caption{The external New York State Bills Dataset} 
\centering 
\begin{tabular}{c c c c c} 
\hline\hline 
Create Date & Bill Title & Bill Subject & Health Area & ...\\ [0.5ex] 
\hline 
2012-05-02 & Prohibits the sale of sugary drinks... & Health & Food Establishment & ...\\ 
2012-10-16 & Concentration of fluoride in water... & Health & Water Quality & ... \\
2015-05-20 & Textbook Transparency Act & Education & N/A & ...\\ [1ex] 
\hline 
\end{tabular}
\label{table:nonlin} 
\end{table}

\subsection{Data Preprocessing \& Cleaning}
For \textit{311\_service\_requests}, we first randomly subsample 30k entries to facilitate further analysis. We are mostly interested in the type of complaint because its granularity is appropriate to map to bills. We consider the top 13 most frequent types of complaints since the rest of them appear less than 0.5 percent throughout the dataset. The selected types of complaints, which reflects health related problems are reported in Table \ref{table:selected_health_problems}.

\begin{table}[ht]
\caption{Selected Health Related Problem Categories.} 
\centering 
\begin{tabular}{c c c c} 
\hline\hline 
Category & Frequency (\%) \\ [0.5ex] 
\hline 
Water System & 34.29\\
Dirty Conditions & 18.46\\
Sanitation Condition & 15.75\\
Rodent & 13.84 \\
Food Establishment & 4.31\\
Air Quality & 3.94\\
Indoor Air Quality & 2.44\\
Food Poisoning & 1.61\\
Hazardous Materials & 1.43\\
Asbestos & 1.06\\
Smoking & 0.98\\
Drinking & 0.70\\
Water Quality & 0.63\\
\hline 
\end{tabular}
\label{table:selected_health_problems} 
\end{table}

In order to analyze the trend of health related problems over time, we bin the data into months and count how many service requests we get per each category. Then we perform a correlation analysis between features to see if some similar categories shall be collapsed into a single category. The results are shown in Figure \ref{fig:correlation_between_categories}. Since \textit{Food Establishment}, \textit{Sanitation Condition} and \textit{Rodent} exhibit a strong correlation between each other, we group them together in the following analyses.

For the \textit{nys\_bills} dataset, since the bill subject is not informative enough, we parse the bill tile to determine the health problem category each bill aims to target. \footnote{For simplicity we use a manually constructed keyword dictionary that aims to trade-off precision over recall, but more complicated techniques like contextual word embeddings might provide a better trade-off.} Finally we also bin the data into months.

\section{Analysis}
\subsection{Change Point Analysis}

\begin{figure}[t]
    \centering
    \includegraphics[width=1.2\linewidth, center]{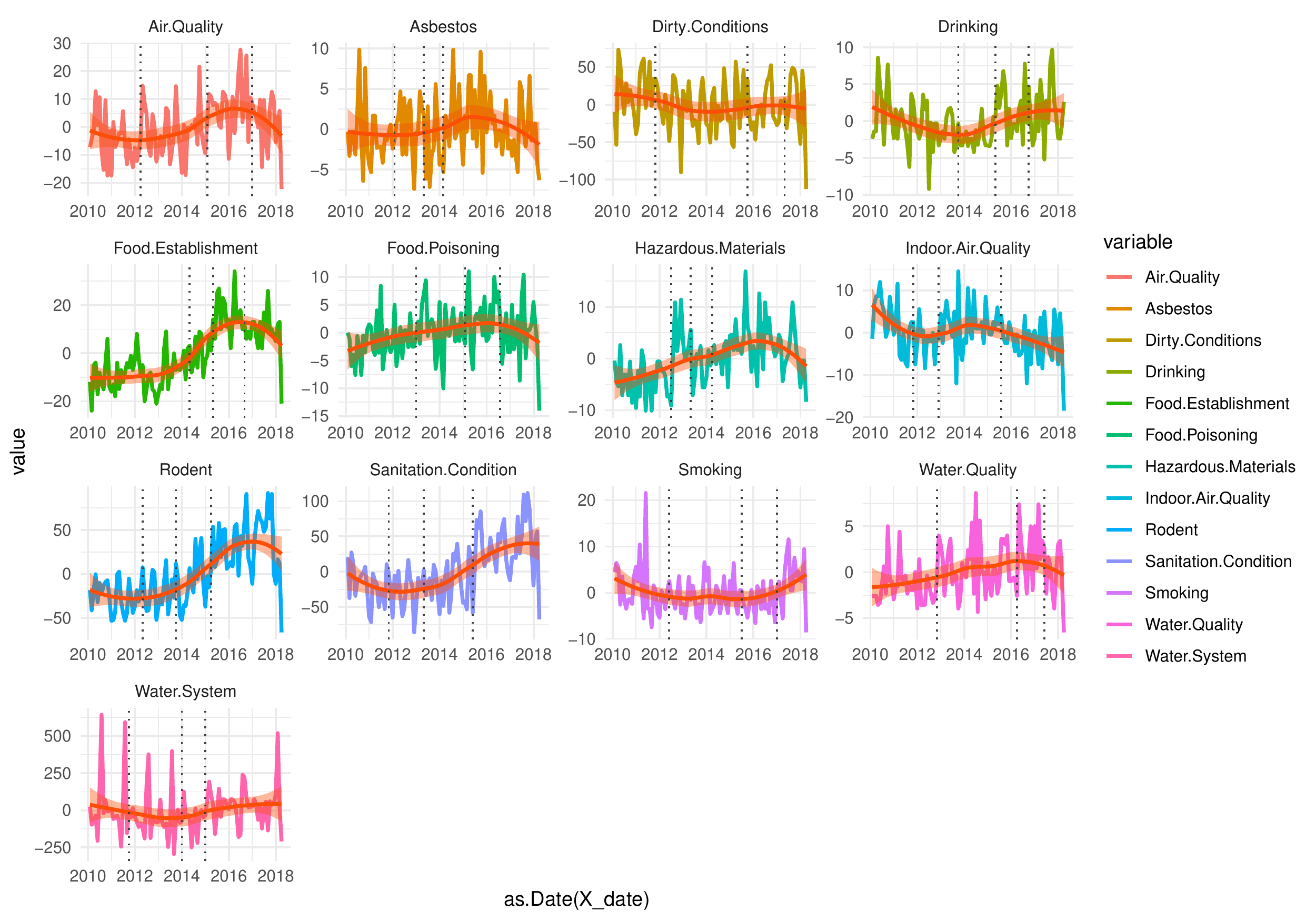}
    \caption{Service Request frequency by Health Area as a proxy for outcome trends from 2010-2018. Change points, where the trend hits an inflection point, are shown in dashed lines. We denote positive legislation as bills coinciding with negative inflection points (service request rate begins to fall) and negative legislation as bills coinciding with positive inflection points.}
    \label{fig:change_point}
\end{figure}

As state legislatures are trying to propose and implement bills to affect health related problems, it's important to analyze the trend of reported health problems, especially at which time points there's a shift in the trend, which might reflect a new bill in effect. To that end, we apply change point analysis, which is also a central research area in time series to a wide range of applications. Detecting critical change points and segmenting time series data into different regimes with different data generating process would be very helpful to reasoning and decision making.


Our main goal here is to detect abrupt change points in people's health related living conditions over the past 8 years in the state of New York, reflected by the number of complaints in aforementioned multiple categories. And we then test the correlation of the change point with relative legislation bills, to find out whether these bills have a significant impact on improving people's living conditions.

We applied the offline version of change point detection [5] onto our processed data. The objective is to minimize a sum of piece-wise loss functions to find the optimal segmentation of a time series, which takes the form below:
\[l(y_{1:T})=\sum_{k=0}^{K}c(y_{t_k:t_{k+1}})\]
where $l(\cdot)$ is the total loss, $c(\cdot)$ is a cost function which measures goodness-of-fit of the sub-series $y_{t_k:t_{k+1}}=\{y_t\}_{t_k+1}^{t_{k+1}}$, and $\{t_1, t_2, \ldots, t_K\}$ is the set of all dividing points, with $t_1=0$ and $t_{K}=T$. We choose the least absolute deviation as our cost, which is a robust estimator of a shift in the central point (mean, median, mode) of a distribution, and takes the following form:
\[c(y_{t_k:t_{k+1}})=\sum_{t_k}^{t_{k+1}}\|y_t - \bar{y}\|_1\]
where $\bar{y}$ is the component-wise median of $y_{t_k:t_{k+1}}$. For the searching algorithm, we used dynamic programming which roughly computes the cost of all sub-sequences of a given time series.

\begin{center}
\begin{figure}[!t]
    \centering
    \includegraphics[width=1.2\linewidth, center]{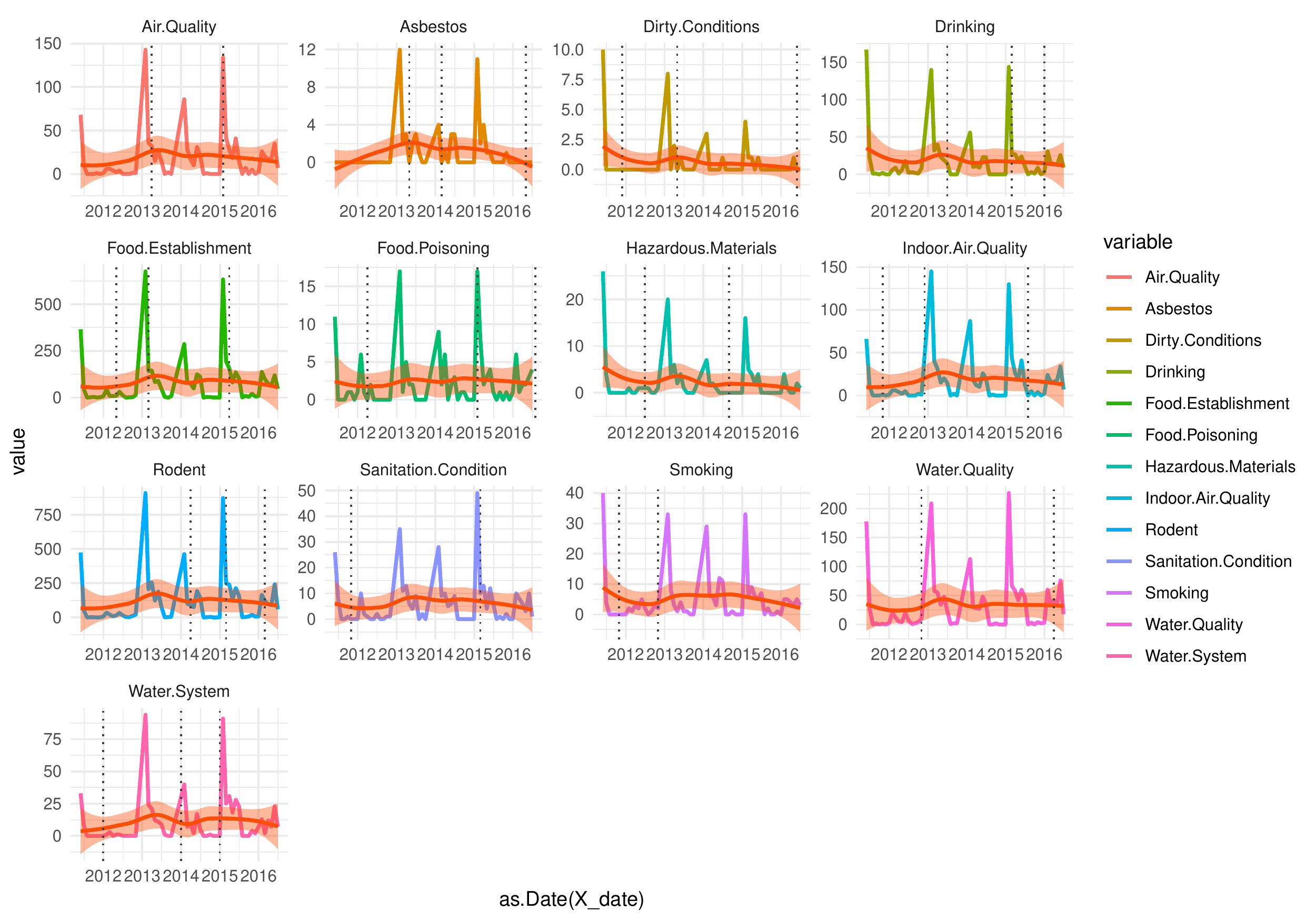}
    \caption{Trend of bills targeting different health problems.}
    \label{fig:change_point_bills}
\end{figure}
\end{center}

\begin{figure}[h]
    \centering
    \includegraphics[width=\linewidth, center]{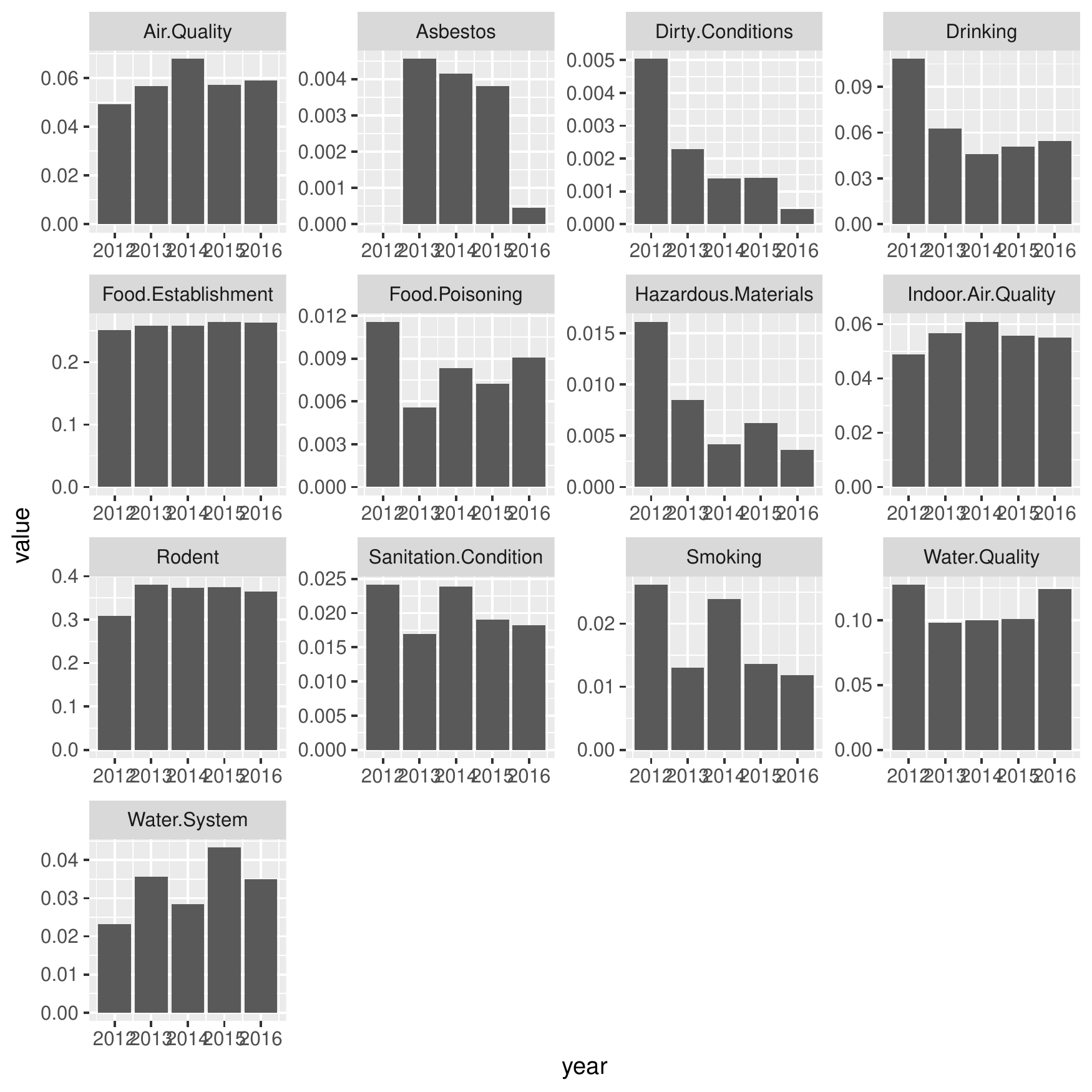}
    \caption{Bar plot}
    \label{fig:bar}
\end{figure}

The results of our change point analysis over the number of complaints in the 13 categories are shown in Figure \ref{fig:change_point}. Note that we observed strong \textbf{seasonal effects} in all of the signals which would potentially hurt the detection, so we removed the seasonal effects first before doing the change detection.

\subsection{Correlation between bills and health problems}

\paragraph{The overall trend of bills}
Similarly, we plot the percentage of bills versus time in Figure \ref{fig:change_point_bills}. We use percentage rather then the absolute numbers because the number of passed bills remain relatively stable over time, and we think percentage reflects the attention paid to different problems. We observe strong seasonality as expected, and the curves mostly exhibit the same trends as the health problem curves, which is unsurprising since more bills might lead to improvement of health conditions, or inversely, more severe problems get more attention from the state legislature.

\paragraph{Do bills fix health problems}
In order to examine whether bills of a certain category fixed their target health problems or not, we perform a chi-squared test to see if there is significant difference in number of bills per year (see Figure \ref{fig:bar}). To correlate the bills with change points, we found in hazardous material, there is significant correlation between number of bills and change points (pval <0.05)

\section{Conclusions \& Future Work}
Our study indicates that there are certain valuable characteristics of the Hazardous Materials legislation that caused a downturn in 311 service requests that made them capable of producing actual health outcomes and the lack of those same characteristics caused the failure of negative legislation to produce better health outcomes. 

The statistical analysis that surfaces these empirically-verifiable distinctions is an essential step in producing future legislation that \textit{learns from past experience}. But it lies for future work, likely in the area of in-depth public policy analysis focused on the specific negative legislation bills, to determine the exact socioeconomic, demographic, political, or implementation factors that resulted in policy failure.

\section*{References}

\medskip

\small

[1] King, Kevin. “State Legislatures Vs. Congress: Which Is More Productive?” Quorum, 2018, {\it www.quorum.us/data-driven-insights/state-legislatures-versus-congress-which-is-more-productive/176/.}

[2] Moore, Carter. “How Long Does It Take to Pass a Bill in the US?” Quora, 2015, {\it www.quora.com/How-long-does-it-take-to-pass-a-bill-in-the-US}.

[3] “How Long Does It Take To Pass And Enact Bills? | PMG.” Parliamentary Monitoring Group, 2015, {\it pmg.org.za/page/How\%20long}.

[4] Gibson, TB (2015) Analyzing the effect of state legislation on health care utilization for children with concussion. {\it Journal of AMA Pediatrics}

[5] Coppin, P., Jonckheere, I., Nackaerts, K., Muys, B. and Lambin, E., 2004. Review ArticleDigital change detection methods in ecosystem monitoring: a review. International journal of remote sensing, 25(9), pp.1565-1596.

\end{document}